\documentclass[%
superscriptaddress,
preprint,
 amsmath,amssymb,
 aps,
floatfix,
]{revtex4-2}
\usepackage{natbib}
\usepackage{graphicx}
\usepackage{latexsym}
\usepackage{amsmath,amssymb}
\usepackage{dcolumn}
\usepackage{bm}

\usepackage{algorithmic}
\usepackage{algorithm}

\begin{document}
\preprint{AON/050511}

\title{Detection of Non-uniformity in Parameters for Magnetic Domain Pattern Generation by Machine Learning}

\author{Naoya Mamada}
\email{mamada.n.aa@m.titech.ac.jp}
\affiliation{Department of Computer Science, Tokyo Institute of Technology, Yokohama 226-8503, Japan}

\author{Masaichiro Mizumaki}%
\email{mizumaki@kumamoto-u.ac.jp}
\affiliation{%
Faculty of Science, Kumamoto University, Kumamoto 860-8555, Japan
}%

\author{Ichiro Akai}%
\email{iakai@kumamoto-u.ac.jp}
\affiliation{%
Institute of Industrial Nanomaterials, Kumamoto University, Kumamoto 860-8555, Japan
}%

\author{Toru Aonishi}%
\email{aonishi@dis.titech.ac.jp}
\affiliation{Department of Computer Science, Tokyo Institute of Technology, Yokohama 226-8503, Japan}

\date{\today}

\begin{abstract}
We estimate the spatial distribution of heterogeneous physical parameters involved in the formation of magnetic domain patterns of polycrystalline thin films by using convolutional neural networks. We propose a method to obtain a spatial map of physical parameters by estimating the parameters from patterns within a small subregion window of the full magnetic domain and subsequently shifting this window. To enhance the accuracy of parameter estimation in such subregions, we employ large-scale models utilized for natural image classification and exploit the benefits of pretraining. Using a model with high estimation accuracy on these subregions, we conduct inference on simulation data featuring spatially varying parameters and demonstrate the capability to detect such parameter variations.
\end{abstract}

\maketitle

\section{Introduction}\label{introduction}
Magnetic thin films are utilized in recording\cite{magnetic_media1, magnetic_media2} media such as magnetic tapes and hard disks, as well as in optical insulators\cite{murayama2006nanoscale, 502265} and switching elements\cite{albert2000spin, sato2001fecobn}. Advancements in these technologies can be expected by controlling the properties of magnetic thin films. However, various factors are associated with these properties, including the type and ratio of elements, temperature, and pressure during the manufacturing process. It is difficult to determine the conditions producing magnetic thin films with the desired properties, and hence, progress has relied on the experimenter's empirical rules and trial and error.

Against this background, research aimed at using machine learning to assist in the creation of magnetic thin films with desired properties has been progressing. For instance, methods have been tried that directly optimize the fabrication conditions using Bayesian optimization\cite{doi:10.1063/1.5123019}\cite{doi:10.1080/27660400.2022.2094698} or reinforcement learning\cite{Tran:22}, and there are techniques that accelerate screening by estimating difficult-to-measure properties of thin films from relatively easily measurable features, such as material element ratios\cite{article331851690} and demagnetization curves\cite{DENGINA2022114797}.

In recent years, attempts have been made to estimate difficult-to-measure physical parameters of thin films from magnetic domain pattern images. Various methods have been employed for this purpose, including topological data analysis \cite{article_persistenthomology}, convolutional neural networks (CNN) \cite{article_348428914, 10.1038/s41524-020-00485-2}, and statistical metrics based on human visual cognition \cite{doi:10.7566/JPSJ.90.044705}. These methods have been used to estimate physical parameters related to magnetic domain pattern generation and have enabled quantitative evaluations of labyrinthine and island structures. Specifically, the use of CNNs has yielded notable results. For instance, our research group \cite{article_348428914} successfully estimated the anisotropy parameter in the time-dependent Ginzburg-Landau equation, while Kawaguchi et al. \cite{10.1038/s41524-020-00485-2} estimated the Dzyaloshinskii-Moriya parameter and magnetic anisotropy dispersion in the Landau-Lifshitz-Gilbert equation from simulation data. Additionally, Kawaguchi et al. successed in estimating the Dzyaloshinskii-Moriya parameter from actual data.

Investigations utilizing machine learning to estimate parameters of magnetic domain patterns typically employ artificial magnetic domain patterns synthesized by micromagnetic simulations, as training data for machine learning models. Micromagnetic simulations reproduce the microscale behaviors of magnetic materials. Factors at the mesoscale, such as crystal size, composition at the grain boundaries, and particle orientation, have been noted to exert a tangible influence upon macroscopic magnetism, rendering micromagnetic simulations a potentially vital contribution to the comprehension and development of magnetic materials. Furthermore, given the existence of real samples where these mesostructures are not spatially homogeneous, there is significance in estimating parameters that are spatially dependent. The magnetic domain patterns synthesized via micromagnetic simulations has been confirmed to resemble those within actual materials possessing analogous parameters\cite{Nakatani_1989}. Additionally, micromagnetic simulations are capable of replicating memory effects and topological melting effects in the formation process of real magnetic domain patterns \cite{PhysRevE.70.046204}, thereby providing a compelling rationale for considering magnetic domain patterns synthesized by micromagnetic simulations as meaningful training data for machine learning models.

In this study, we estimate the spatial distribution of non-uniform physical parameters for the analysis of polycrystalline thin films by using a CNN. For the entire magnetic domain pattern, we estimated the physical parameters from patterns within a small subregion of the window and estimate the spatial distribution of physical parameters by shifting this window. Our previous research showed that the accuracy of parameter estimation deteriorates when the pattern area is small. Here, we significantly improve the estimation accuracy of the estimation from small subregion patterns compared with the preceding model and achieved the above objective through the following methods:
\begin{itemize}
\item \textbf{Large-scale models} - Large-scale models have many model parameters. Although they require larger computational resources and are more prone to overfitting when training data is scarce, they can handle more complex data if sufficient training data is available.
\item \textbf{Pretraining} - Pretraining\cite{pan2009survey, 9134370}, also referred to as transfer learning, involves initially training the model on data from a different domain before training it on the target domain data. This accelerates learning and prevents overfitting for models with many parameters. Pretraining is particularly useful when the cost of obtaining target domain data is high. Pretraining is often done on ImageNet\cite{deng2009imagenet}, which is a dataset of natural images used for image classification. Pretraining with ImageNet has also been applied in medical research, such as for CT\cite{doi:10.1148/ryai.2019180066} and ultrasound\cite{cancers13040738} images, as well as in material science for predicting the physical properties of magnets \cite{8955506} and classifying defects in metal materials \cite{pretrained_cnn_defects}.
\end{itemize}

The previous research by Kawaguchi et al.\cite{10.1038/s41524-020-00485-2} and our own work\cite{article_348428914} used small-scale CNN models designed without pretraining. In this study, we examined the usefulness of large-scale CNN models and pretraining in predicting the properties of magnetic thin films and obtained positive results. Our contributions are as follows:

\begin{itemize}
\item We performed inference on test data with spatially varying parameters and demonstrated that we could detect changes in those parameters.
\item We showed that large-scale models, which are designed for natural image classification, are effective for estimating small-region simulation parameters, and that pretraining with natural images is useful.
\end{itemize}

This paper is organized as follows. Sect.\ref{numerical experiment} gives an overview of teh numerical experiments. Sect.\ref{SimulationandMagneticDomainPatternAcquisition} explains the simulation model used to generate the magnetic domain pattern data.  Sect.\ref{training} describes the machine-learning model that we developed to analyze the magnetic domain patterns. Sect \ref{results} evaluates the effectiveness of our model by verifying if the learning model can correctly infer the parameter changes. Sect.\ref{discussion} presents the results and discussion of each numerical experiment. Sect.\ref{conclusion} concludes this paper.

\section{numerical experiments}\label{numerical experiment}
\subsection{Overview of Numerical Experiments}\label{Problem setting}
The numerical experiments conducted in this study are composed of two sequential experiments, as depicted in Fig. 1. Experiment I utilizes data obtained from simulations that were executed with spatially uniform parameters, and it aims to train the model to estimate parameters, noise coefficient $D$ and the exchange interaciton coefficient $J$ described in Sect. IIB. 
Experiment II employs the machine learning model trained in Experiment I. The machine learning model infers the spatial distribution of $D$ and $J$ based on data from simulations in which $D$ was spatially varied while $J$ was fixed. Experiment II aims to investigate wheter the machine learning model can detect spatial change in simulation parameter $D$ and, even in the presence of such spatial variations in parameter $D$, wheter it can reliably infer the fixed parameter $J$.

\begin{figure}[h]\label{fig:flowchart}
\includegraphics[width=\linewidth]{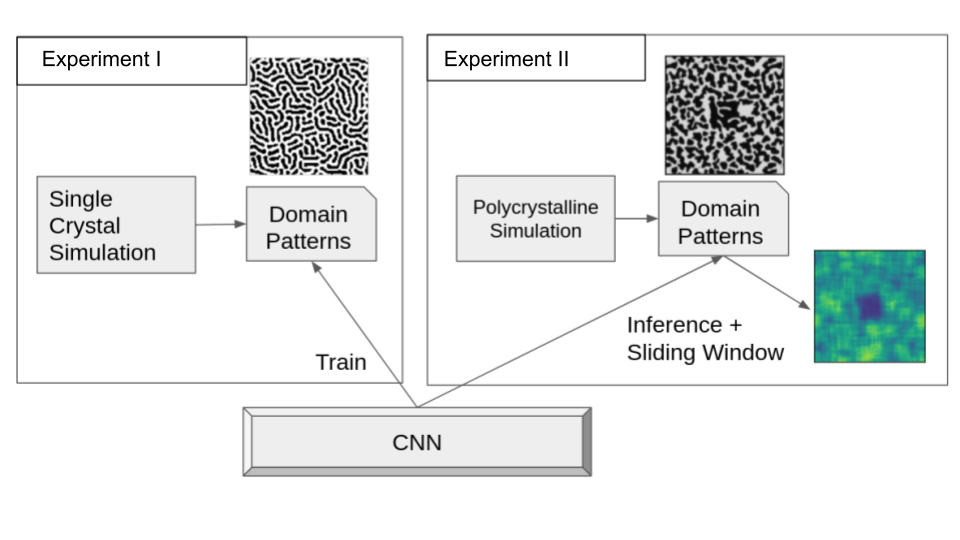}
\caption{\textbf{Flow Chart of Our Numerical Experiment}\\ 
The flow chart depicts the outline of two sequential experiments. Experiment I is for training and model selection. Experiment II is for inference of spatial distribution of parameters with the machine learning model trained in Experiment I.}
\end{figure}

\subsection{Simulation and Magnetic Domain Pattern Acquisition}\label{SimulationandMagneticDomainPatternAcquisition}
We modeled the average value of the $z$-component of the spin in a small area at $\mathbf{r} \in \mathbb{R}^2$ on a magnetic thin film placed on the xy plane as $m_z(\mathbf{r}) \in [-1, 1]$ and simulated the hysteresis process of the thin film by using time dependent Ginzburg-Landau\cite{10.1103/physrevb.72.094406} equation
\begin{equation}
  \label{llg} \begin{split}
  \frac{\partial m_z}{\partial t} &= (1-m_z^2) \left( h(t) + A_0 (1 + D \eta(\mathbf{r}))m_z - \gamma \int d\mathbf{r}' \frac{m_z(\mathbf{r}')}{|\mathbf{r}-\mathbf{r}'|^3} \right) \\ &+ J \nabla^2 m_z
  \end{split}
\end{equation}
Here, $h(t)$ is the time-dependent external magnetic field, $A_0$ is the magnetic anisotropy coefficient, $\eta$ is spatially independent and time-invariant noise following a standard normal distribution, $D$ is the noise coefficient, $\gamma$ is the dipole interaction coefficient, and $J$ is the exchange interaction coefficient. We initialized $m_z$ with random values from a uniform distribution $[-0.05, 0.05]$. The time step width was set to 0.1.\newline
The external magnetic field $h$ was varied according to the following equation.
\begin{align}
\label{external}
h(t) = 
  \begin{cases}
  10^{-5}Ht &(t < 10^5)\\
  H(2 - 10^{-5}t) &(10^5 \leq t \leq 3 \times 10^5)\\
  H(10^{-5}t - 4) &(3 \times 10^5 < t \leq 5 \times 10^5)\\
  \end{cases}
\end{align}

Here, $H$ is the maximum value of the external magnetic field during the hysteresis process. In this setting, the change in the external magnetic field $h$ is very slow compared with the change in Eq. 1.; thus, Eq. 1 can always be considered as representing a stationary state. Under the conditions, the hysteresis process shows a closed hysteresis loop, as shown in Fig. 2.

\subsubsection{Generation of spatially uniform data}\label{Generation of spatially uniform data}
In the simulation, the space was discretized into $256 \times 256$ pixels, and periodic conditions were set. To obtain a variety of hysteresis processes, we conducted simulations with different values of $D$ and $J$. We fixed the other parameters $A_0 = 1.0, \gamma = 0.095J$, and $H = 0.3$, as in the previous research\cite{10.1103/physrevb.72.094406}.  
We varied both $D$ and $J$ in the range from 0.1 to 2.0 in steps of 0.1. When both $D$ and $J$ were small, the calculation became unstable, so we excluded such combinations from the datasets. The combinations of $D$ and $J$ in the dataset are listed in the Appendix.

In the hysteresis process illustrated in Fig. 2, the magnetic domain pattern utilized for machine learning was acquired at the moment the average value of $m_z$ throughout the entire system attained zero for a second time following its initial time. The values of $m_z$ tend to approximate either +1 or -1. The values of $m_z$ are binarized to take either +1 or -1 with a threshold of 0. Examples of the obtained magnetic domain patterns are shown in Figs. 3 and 4.

\begin{figure}[h]\label{fig:hys_curve}
\includegraphics[width=\linewidth]{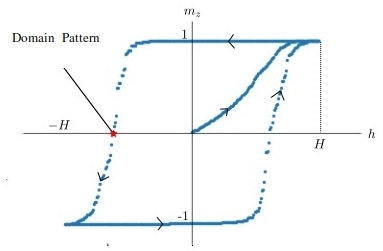}
\caption{\textbf{Plot of the simulated hysteresis process}\\ A representative example of the hysteresis simulation used in this study. The vertical axis represents the average value of magnetization ($m_z$) of the entire system at each time step, and the horizontal axis represents the external magnetic field ($h$). The initial state is at the origin, and the time evolution is shown by the black arrows. The magnetic domain pattern at the point where the average magnetization becomes zero for the second time, indicated by the red star, was used to estimate the simulation parameters.}
\end{figure}

\begin{figure}[h]\label{fig:sample_patterns}
\includegraphics[width=\linewidth]{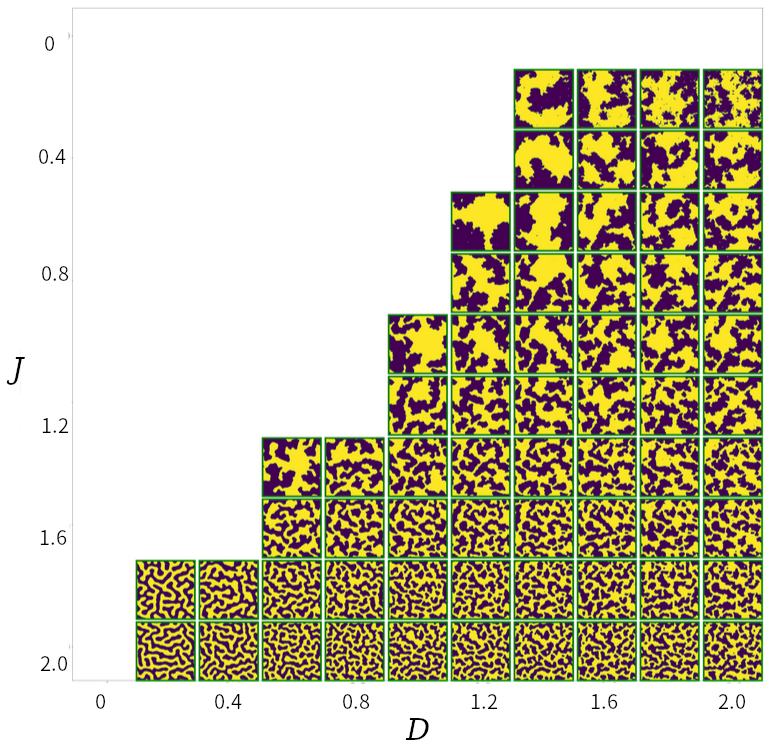}
\caption{\textbf{Examples of magnetic domain patterns with fixed initial values}\ Magnetic domain patterns generated using the same initial values and different combinations of $D$ and $J$ are shown.}
\end{figure}

\begin{figure}[h]\label{fig:patterns_by_seeds}
\includegraphics[width=\linewidth]{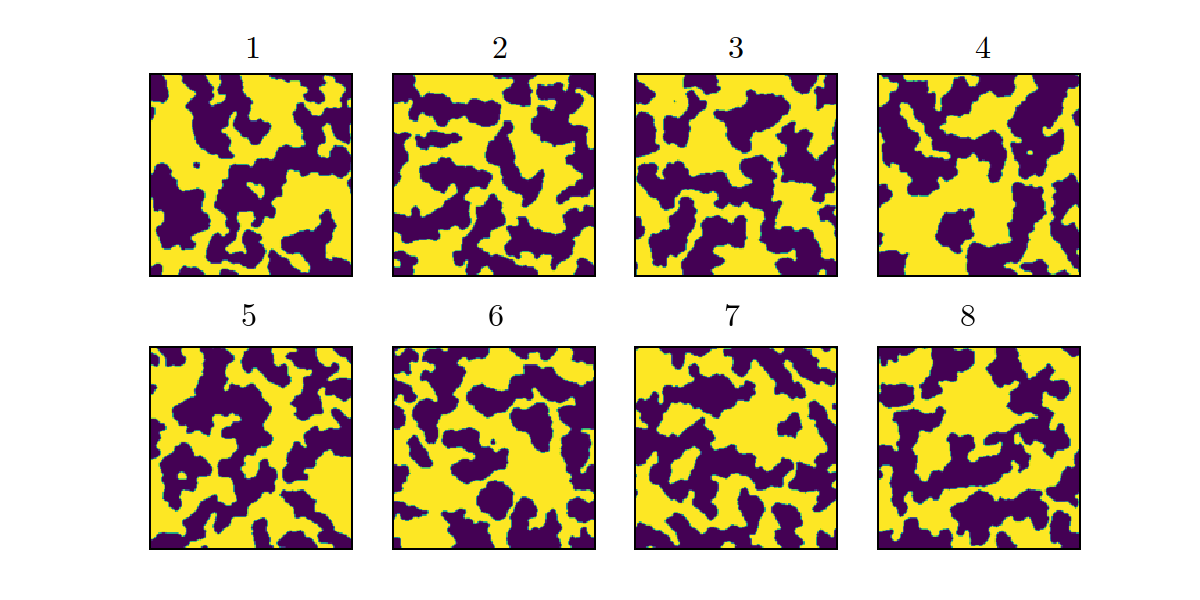}
\caption{\textbf{Examples of magnetic domain patterns with fixed values other than the initial ones}\ Magnetic domain patterns generated with $D=J=1.5$ and different random seed initial values are shown.}
\end{figure}

We performed 12 simulations with random seeds $s$ for the initial values, where $s \in \{1, ..., 12\}$, for each combination of $D$ and $J$ and created a dataset.

\subsubsection{Generation of Spatially Inhomogeneous Data}\label{Generation of Spatially Inhomogeneous Data}

We conducted an numerical experiment to confirm whether the model learned with the data collected in \ref{SimulationandMagneticDomainPatternAcquisition} could detect changes in the parameter $D$ from domain patterns generated from simulations with spatially varying $D$. We discretized the simulation space into $512\times 512$ pixels, set periodic boundary conditions, and called the region within 192 pixels from the boundary the \textbf{outer}, and the rest the \textbf{inner}. We fixed the value of $D$ in the \textbf{outer} at 1.0 and changed the value of $D$ in the \textbf{inner} from 0.1 to 1.9 in increments of 0.3, and performed the hysteresis process simulations. We refer to the value of $D$ in the \textbf{inner}  $D_\text{in}$. The other parameters were fixed $A_0=1.0,\gamma=0.095J, H=0.3, J=2.0$. Fig. 5 shows a schematic diagram of the spatial variation in $D$.

As described in Sect.\ref{Generation of spatially uniform data}, we obtained magnetic domain patterns in the hysteresis process with an average magnetization of zero under these conditions. Examples of these magnetic domain patterns are shown in Fig. 6.

\begin{figure}[h]\label{fig:changing_D}
\centering
\includegraphics[scale=0.45]{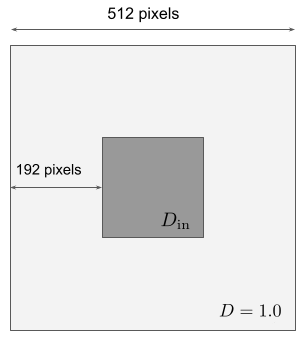}
\caption{\textbf{Schematic diagram of the spatial variation of the parameter} \newline
Magnetic domain patterns were generated in the simulation by varying the value of $D$ in the \textbf{inner} region ($D_\text{in}$) and fixing the value in the \textbf{outer} region to $D=1.0$. We determined whether the value of $D$ can be estimated from the magnetic domain patterns within a sliding window defined by Algorithm 1 and whether this change in $D$ could be detected.}
\end{figure}

\begin{figure}[h]\label{fig:changing_D2}
\centering
\includegraphics[width=\linewidth]{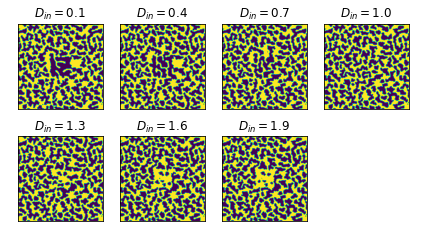}
\caption{\textbf{Sample data of spatially inhomogeneous parameters}\newline
Examples of magnetic domain patterns found by varying $D_\text{in}$ in the \textbf{inner} region and setting $D=1.0$ in the \textbf{outer} region. The image's side length is 512 pixels.}
\end{figure}

\subsubsection{Computational Environment}
The simulation program was written in Fortran. The generation of spatially homonegeous data as described in Sect.\ref{Generation of spatially uniform data} required approximately 10 hours per pattern on a signle core of AMD Ryzen 3950X CPU. Moreover, for generating spatially the inhomogeneous data as outlined in Sect.\ref{Generation of Spatially Inhomogeneous Data}, the spatial resolution was increased from the 256 $\times$ 256 pixels used in Sect.\ref{Generation of spatially uniform data} to 512 $\times$ 512 pixels; this took roughly 40 hours per pattern.
Approximately 3,000 magnetic domain patterns were needed, so a computational cluster consisting of 17 nodes was constructed. Batch job management was performed using the Slurm Workload Manager \cite{10.1007/10968987_3}. Computation of all patterns took approximately 10 days in total.

\subsection{Machine Learning}
\subsubsection{Structure of Learning Model}\label{Structure of Learning Model}
 We developed two CNNs to simultaneously estimate $D$ and $J$. The first was a small-scale CNN model (Model0-6, \textbf{mod}) based on our previous research\cite{article_348428914}. The second was a large-scale CNN model (\textbf{reg}) based on the natural image classification model RegNetX-16GF\cite{https://doi.org/10.48550/arxiv.2101.00590}. The models shared core components, including convolutional layers, batch Normalization\cite{DBLP:journals/corr/IoffeS15}, and ReLU activation functions\cite{agarap2018deep}. Model0-6 consisted of seven convolutional layers and a maximum of 36 channels, while RegNetX-16GF comprised 60 convolutional layers with a maximum of 912 channels and incorporated techniques such as skip connection\cite{He2016DeepRL} and group convolution\cite{NIPS2012_c399862d}.

Table I presents the number of parameters for each model and their respective accuracy at natural image (ImageNet) classification. Model0-6 had a significantly fewer parameters, approximately $\frac{4}{10000}$ compared with RegNetX-16GF, which rendered it unable to cope with the complexity of ImageNet. Consequently, its accuracy was 73.3\% lower in comparison to RegNetX-16GF.

\begin{table}[h]
\centering
\caption{Number of parameters and accuracies of each model}
\label{table:models_info}
\begin{tabular}{lcc}

         & Number of Parameters            & ImageNet Accuracy (\%) \\ \hline
Model0-6 & 16000             & 6.2                   \\
RegNetX-16GF   & 3.92$\times 10^7$ & 80.1              
\end{tabular}
\end{table}

We also conducted numerical experiments with RegNetX-16GF using the ImageNet pretraining (\textbf{reg\_pret}). For Model0-6, there were no significant difference in $D$ and $J$ in terms of learning and inference with or without pretraining, so we only considered the case without pretraining.

To make a regression model, the original model's penultimate layer was replaced with a layer containing only two neurons for estimating $D$ and $J$. We implemented the models using the deep learning library PyTorch\cite{NEURIPS2019_9015}.

We used input images consisting $64 \times 64$ pixels. In our previous research\cite{article_348428914}, we evaluated the relationship between input size and parameter estimation accuracy for Model0-6's input magnetic domain pattern images. The estimation accuracy for $64 \times 64$ pixel input images was found to decrease by 60\% compared with the accuracy of $256 \times 256$ pixel images halving 16 times the resolution.

In addition to RegNet, we conducted numerical experiments with several well-known deep-learning models, including MobilenetV3-Large\cite{mobilenetv3}, EfficientNet-B3\cite{efficientnet}, MnasNet1-0\cite{mnasnet}, and ResNext50\cite{resnext}. RegNet achieved the highest accuracy, so we conducted a detailed examination of it. The numerical experimental results for these deep-learning models are shown in Appendix B.
The pretrained weights of the deep learning models, other than Model0-6, were downloaded from Torchvision\cite{torchvision2016} codebase. RegNet model's pretraining procedure following Torchvision's training recipe\cite{training_recipe} was as follows.

The model was trained on ImageNet dataset for 100 epochs with a batch size of 64, a weight decay of 0.00005, and an initial learning rate of 0.4. The learning rate was scheduled according to a cosine annealing schedule\cite{DBLP:journals/corr/LoshchilovH16a}, with a linear warmup over 5 epochs and a post-warmup decay of 0.1. The training is distributed over multiple processes, with 8 processes per node. The \textbf{mod} model was trained following the same recipe described above, where we used single node and single process. Training recipes for other models are also in public on Torchvision codebase.

\subsubsection{Experiment I: Training and testing with uniform data}\label{training}
The training data were the spatially uniform data described in Sect.IIB1. $\text{N}_\text{data}$ number of the random seeds were randomly selected from $\{3, 4, ..., 12\}$, and images generated with the selected seeds were used in training. For training, as described above the data were prepared for each combination of $J$ and $D$ shown in Table III of Appendix A. During training,
data augmentation was employed by randomly cropping $64 \times 64$ pixel patches from the original $256 \times 256$ pixel images and applying random rotations of $k \in \{0, 90, 180, 270\}$ degrees. The mean absolute error was used as the loss function, and the optimization was executed with momentum stochastic gradient descent, where the initial learning rate was set to 0.001, and was gradually reduced by a factor of 0.8 every 500 steps. Training was terminated when the validation error with a patch of the seed $s = 2$ failed to decrease for three consecutive epochs or when 100 epochs had elapsed. Then, the trained model was tested with the central $64 \times 64$ pixels of the image of the seed $s=1$ which was rotated with the four angles of 0, 90, 180 and 270 degrees. The final test error was calculated as the average of the four prediction errors obtained from these rotation images. This random seeds selection, training, and testing procedure was repeated 20 times for each combination of model \{\textbf{mod}, \textbf{reg}, \textbf{reg\_pret}\} and $\text{N}_\text{data}$ $\in$ \{1, 4\}.

\subsubsection{Experiment II: Inference of nonuniform parameter}\label{inference}
In Experiment II, the distributions of $D$ and $J$ was estimated from the spatially non-uniform data described in Sect.IIB2. The trained models in Experiment I were employed in the sliding window approach given in Algorithm 1. This approach constructs the estimation map of $D$ and $J$ by shifting the $64 \times 64$ pixel window one pixel at a time, cropping a $64 \times 64$ pixel patch from the target image by the window and inferring the values of $D$ and $J$ from the patch. When constructing the estimation map using the sliding window approach on one individual images with 512 × 512 pixels, the inference was repeated approximately 250 thousands times ($512 \times 512$). This calculation using a single NVIDIA RTX A4000 Graphical Processing Unit, necessitated an approximate duration of 500 seconds.

\begin{figure}[h]
\begin{algorithm}[H]
    \caption{\textbf{Inference using Sliding Window} \newline $A, P$ are $512 \times 512$ pixel arrays, where $A$ represents the input magnetic domain pattern and $P$ stores the inference results. The colon symbol denotes array slicing, and $A[i-32: i+31, j-32: j+31]$ is a $64 \times 64$ pixel sub-region extracted from the $i-32$th row to the $i+31$st row and the $j-32$th column to the $j+31$st column of $A$. Since the simulation had periodic boundary conditions, the periodicity in the column direction of A is introduced as \newline $A[i, j] = 
    \begin{cases}
    A[i + 512, j] & (i \leq 0)\\
    A[i - 512, j] & (513 \leq i) \\
    A[i, j] & (\text{otherwise})
    \end{cases}$ \newline The periodicity in the row direction of A is also introduced as the same manner. $\text{MODEL}: \{-1, 1\}^{64 \times 64} \mapsto \mathbb{R}$ represents the inference model.}
    \label{alg1}
    \begin{algorithmic}[]
        \FOR{$0 < i < 513$}
            \FOR{$0 < j < 513$} 
                \STATE{P[i, j] = MODEL(A[i-32: i+31, j-32: j+31]) \newline j = j + 1}
            \ENDFOR
            \newline
        i = i + 1
    \ENDFOR
    \end{algorithmic}
\end{algorithm}
\end{figure}

\section{Results}\label{results}
\subsection{Experiment I: Parameter Estimation for Spatially Uniform Data}
Here, we conducted training and testing of the models, \textbf{reg}, \textbf{reg\_pret} and \textbf{mod} for spatially uniform data described in Sect.IIA1. As explained in Sect. IIC2, the training and testing procedure was repeated 20 times for each combination of model {mod, reg, reg\_pret} and $N_\text{data} \in \{1, 4\}$, and 20 test errors for each combination were evaluated as follows.

\begin{figure}[h]\label{fig:boxplots}
\includegraphics[width=\linewidth]{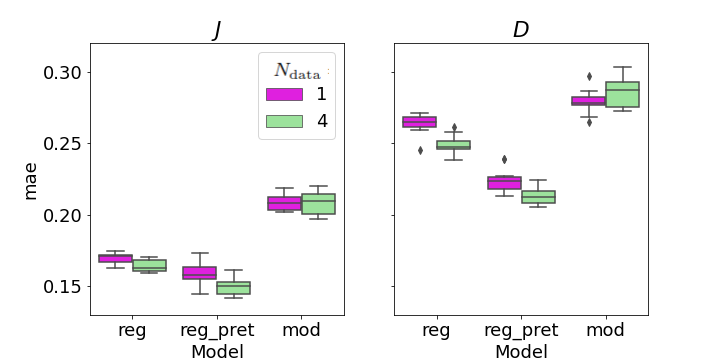}
\caption{\textbf{Mean absolute error for each model}\\
The distribution of the mean absolute error (mae) for \textbf{reg}, \textbf{reg\_pret}, and  \textbf{mod} is shown in boxplots. The comparison is between the cases of $N_\text{data}=1$ and $4$.}
\end{figure}

\begin{table}[h]
\centering
\label{table:error_median}
\caption{\textbf{Median of mean absolute errors for each numerical experimental condition}}
\begin{tabular}{ccccc}
Model                & \multicolumn{4}{c}{target}                    \\
                     & \multicolumn{2}{c}{$J$} & \multicolumn{2}{c}{$D$} \\ \cline{2-5} 
\multicolumn{1}{l}{} & \multicolumn{4}{c}{$\text{N}_{data}$}                     \\
                     & 1         & 4         & 1         & 4         \\ \hline
\textbf{reg}                  & 0.171     & 0.162     & 0.265     & 0.247     \\
\textbf{reg\_pret}            & 0.157     & 0.151     & 0.223     & 0.212     \\
 \textbf{mod}                  & 0.208     & 0.210     & 0.278     & 0.287  
\end{tabular}
\end{table}

\begin{figure*}[t]\label{fig:prediction_boxplot}
\includegraphics[width=\linewidth]{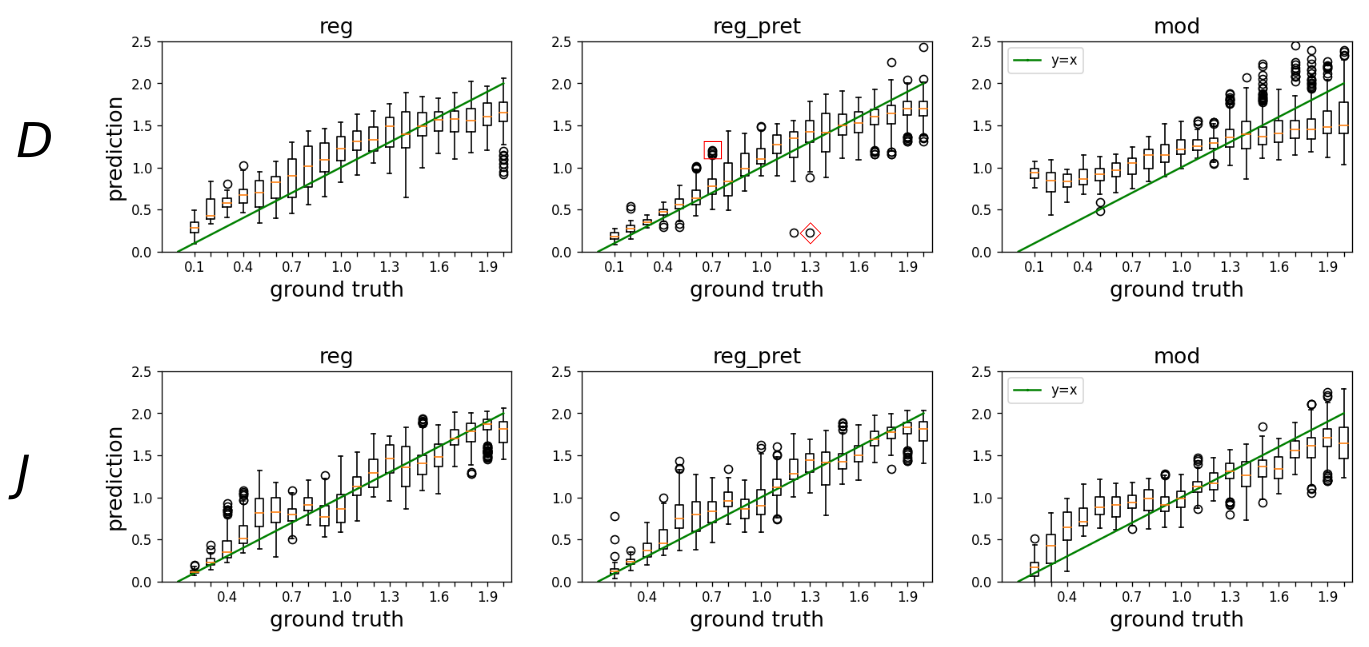}
\caption{\textbf{Relationship between true values and predicted values}\\ The horizontal axis represents the true values of the target variables, and the vertical axis represents the predicted values for each model. The distribution of predicted values with respect to true values is shown in boxplots. The green line is the line $y=x$. The top row and bottom row correspond to the target variables $D$ and $J$, respectively, and the columns from left to right represent \textbf{reg}, \textbf{reg\_pret}, and \textbf{mod}. In the top row, middle column panel, two of the outliers whose true values are $D=0.7$ and $D=1.3$ are marked with red square and diamond.}
\end{figure*}

\begin{figure}[h]\label{fig:outliers}
\includegraphics[width=\linewidth]{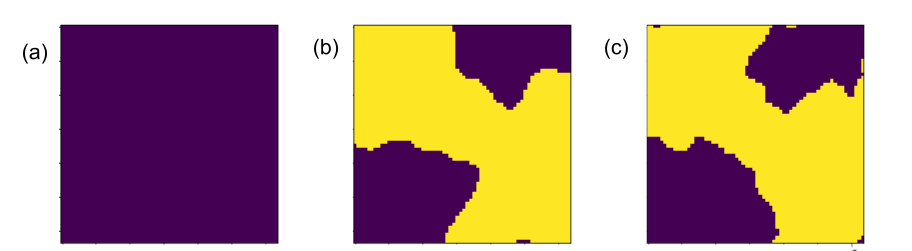}
\caption{\textbf{Patches for Outliers and its Similar Learning Patch}\\ (a): The patch of test data that has truth value $D = 1.3$ and predicted to be around 0.2; corresonds to diamond marker in Fig. 8. (b): The patch of test data that has truth value $D = 0.7$ and predicted to be around 1.3; corresponds to square marker in Fig. 8. (c): The patch of learning data similar to (b) that has truth value $D = 1.3$.}
\end{figure}

\subsubsection{Estimation Accuracies for $D$ and $J$ Across Models}
Fig. 7 displays the distribution of mean absolute errors for the estimation of target variables $J$ and $D$ in each model, and Table II shows the median values of these distributions. When $N_\text{data} = 4$, the \textbf{reg\_pret} model performed the best at estimating both $D$ and $J$, with errors of 0.212 and 0.151, respectively, Regardless of the target variable or $N_\text{data}$, the \textbf{reg\_pret} model produced the smallest errors, while  \textbf{mod} model consistently yielded the largest. Additionally, the error for $D$ was noticeably larger than that for $J$, regardless of the model.

Fig. 8 shows boxplots representing the distribution of predicted values for each model with respect to true values. As indicated in Fig. 8, for $\text{N}_\text{data}=4$, all models were able to infer the correct order of magnitude for the target variables; however, their estimation accuracies differed. The \textbf{reg\_pret} model exhibited the lowest error, while the  \textbf{mod} model performed noticeably worse. In the \textbf{mod} model, when the true value of $D$ was 0.1, the median estimated value was 0.9, and when the true value of $D$ was 2.0, the median estimated value was 1.4. Thus, the relationship between the estimated and true values of $D$ deviated significantly from the green $y=x$ line.

\subsubsection{Comparison of Models with and without Pretraining}
A comparison of the mean absolute errors of the \textbf{reg} and \textbf{reg\_pret} models in Fig. 7 and Table II, makes it clear that the \textbf{reg\_pret} model achieved higher estimation accuracy under all conditions. The greatest improvement in accuracy was observed for the target variable $D$ when $\text{N}_\text{data}=1$, with the median of mean absolute errors decreasing by 0.042 due to pretraining. Even when $\text{N}_\text{data}=4$ in the case of target variable $J$, which had the smallest improvement margin, pretraining led to a 0.011 decrease in the median of the mean absolute errors.

\subsubsection{Comparison of Amounts of Training Data, $\text{N}_\text{data}$}
Comparing the mean absolute errors for $\text{N}_\text{data}=1$ and $4$ in Fig. 7 and Table II, makes it evident that the estimation accuracy was higher for $\text{N}_\text{data}=4$ under all conditions for both \textbf{reg} and \textbf{reg\_pret} models. Therefore, increasing the amount of training data led to an improvement in the accuracy of these models. In contrast, the  \textbf{mod} model exhibited a larger median of mean absolute errors for $\text{N}_\text{data}=4$ compared with $\text{N}_\text{data}=1$, with increases of 0.002 for $J$ 0.009 for $D$.

\subsubsection{Outliers and Related Domain Patterns}
In Fig. 8, the red square and diamond markers denote the outlier estimations for the true $D$ values of 0.7 and 1.3, respectively. Figs. 9 (a) and (b) display $64 \times 64$ pixel patches of test data that led to these outliers, while Fig. 9(c) presents a patch of training data with true $D$ value of 1.3 that looks similar to the patch in Fig. 9(b). Out of twenty reg\_pret models trained using different randomly selected inital random seeds(see Sect. IIC2), one of the twenty trained models estimated the patch in Fig. 9(a) as the outlier, and eight of the twenty trained models estimated the patch in Fig. 9(b) as the outlier.

\begin{figure*}[t]\label{fig:dif_d_preds}
\includegraphics[width=\linewidth]{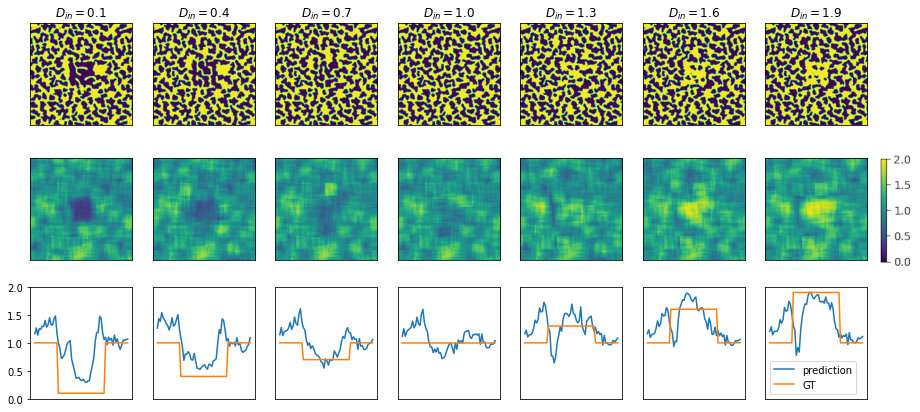}
\caption{\textbf{Estimation maps of $D$ created by sliding window approach}\\Top row: input images which are the same as Fig. 6. Middle row: estimation map of $D$. Bottom row: prediction and ground truth of $D$ along the horizontal line through the center of images.}
\end{figure*}

\begin{figure}[h]\label{fig:different_d_preds_lineplot}
\includegraphics[width=\linewidth]{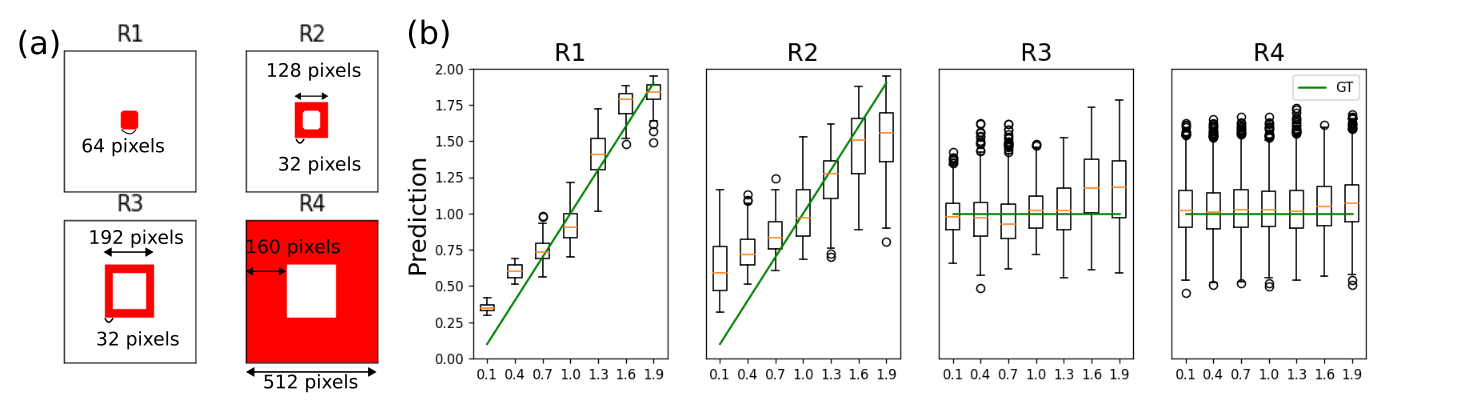}
\caption{\textbf{Distribution of Estimation Values in Separated Regions}\\ (a) The schematics of the separated regions. The \textbf{inner} region is separated into R1 and R2, while the \textbf{outer} region is separated into R3 and R4. (b) The distribution of estimation values with respect to $D_\textbf{in}$ is shown as boxplots. The green lines indicate ground truth values in the separated regions.}
\end{figure}

\subsection{Experiment II: Detection of Variations in Parameters from Spatially Nonuniform Data}\label{Detection of Spatial Variations in Parameters from Magnetic Domain Patterns}
Here, using the sliding window approach described in Sect.IIC3, we estimated the distributions of $D$ and $J$ from the spatially non-uniform data described in Sect.IIA2 as follows.

Fig. 10 shows the magnetic domain patterns generated by the simulations in which $D$ was spatially varied while $J$ was fixed as described in Sect.IIB2, as well as the estimation map of $D$ obtained using and the sliding window approach with the reg\_pret model trained in Experiment I. In the magnetic domain patterns shown in Fig. 10, the value of $D_\text{in}$ in the \textbf{inner} region varies from 0.1 to 1.9, while in the \textbf{outer} region it is fixed at $D = 1.0$. The center region of the estimation map in Fig. 10 shows an increasing trend as $D_\text{in}$ increases.

For the spatially non-uniform data described in Sect.IIA2, the nonuniformity of $D$ values in the image was sought to be detected using the sliding window approach. Fig. 10 shows the magnetic domain patterns generated by the simulations under the heterogeneous conditions for $D$ described in Sect.\ref{Generation of Spatially Inhomogeneous Data}, as well as the map of the estimated values of $D$ obtained using the \textbf{reg\_pret} model and the sliding window method described in Sect.\ref{inference}. In the magnetic domain pattern samples shown in Fig. 10, the value of $D_\text{in}$ in the center region varies from 0.1 to 1.9, while the \textbf{outer} region it is fixed at $D = 1.0$. The center region of the estimated map in Fig. 10 shows an increasing trend for $D_\text{in}$. 

Next, in order to apprise how the inference accuracy is influenced by the boundary between the inner and outer regions, we splitted the inner region into R1 and R2 and the outer region into R3 and R4 as shown in Fig 11(a). Moreover we evaluated the difference in accuracy on the spitted regions depending on the distance from the boundary. 

Fig. 11(b) shows how estimated values are distributed relative to the true value of $D_\text{in}$. In both R1 and R2, as depicted in the figures, the center of the distribution of the estimated values ascends in alignment with the green $y = x$ line, whereas the error on R2 from the true value of $D_\text{in}$ is larger than that of R1. On the other hand, in both R3 and R4, the distribution of inference values are around the green $y = 1$ line, though the error on R3 from the true value of $D$ is larger than that of R1 for $D =1.6$ and $1.9$. Therefore, though the errors on R2 and R4 near the boundary are larger than those of R1 and R4, the inference values on R2 and R4 are distributed around the true values.

\begin{figure}[h!]\label{fig:roc_curve}
\includegraphics[width=\linewidth]{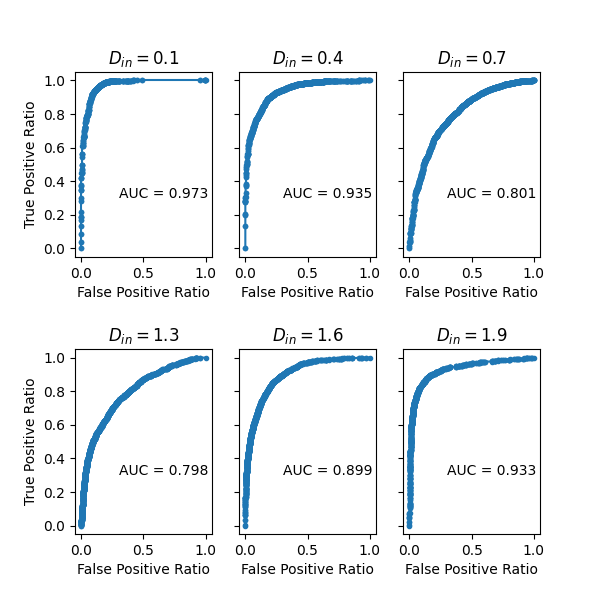}
\caption{\textbf{ROC analysis of distribution of predicted values in inner and outer regions}\\The ROC curve and AUC are shown for the binary classification where the distribution of estimated values in the region with larger true $D$ is the positive instance and that with smaller true $D$ is the negative instance.}
\end{figure}

To quantitatively evaluate the separability of estimated values in the \textbf{inner} and \textbf{outer} regions, we performed a receiver operating characteristic (ROC) analysis as shown in Fig. 12. The ROC curve and AUC are shown for the binary classification where the distributions of estimated values in the region with larger $D$ is the positive instance and that of the estimated values in the region with smaller true $D$ is the negative instance. Fig. 12 indicates that the larger the difference between the \textbf{inner} $D_\text{in}$ and outer $D$, the higher the AUC score becomes. The AUC score was the lowest (0.798) when $D_\text{in} = 1.3$.

\begin{figure*}[t]\label{fig:dif_d_preds_j}
\includegraphics[width=\linewidth]{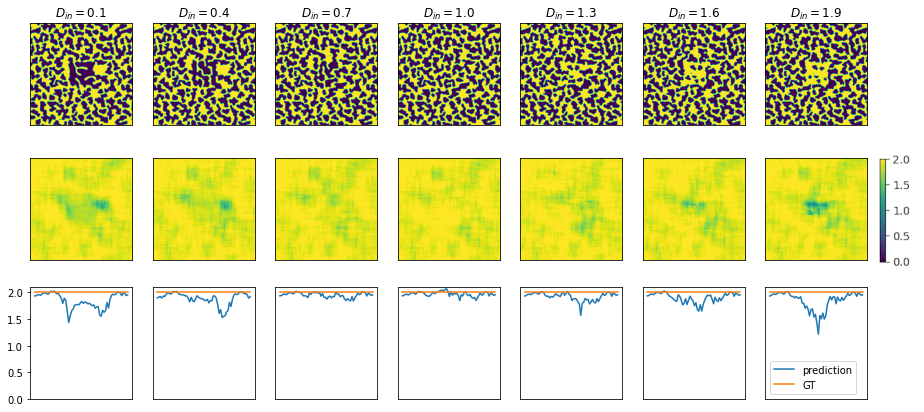}
\caption{\textbf{Estimation maps of $J$ created by sliding window approach}\\Top row: input images. Middle row: estimation map of $J$. Bottom row: prediction and ground truth of $J$ along the horizontal line through the center of images.}
\end{figure*}

As descibed above, the magnetic domain patterns were generated under the condion that $D$ was spatially varied while $J$ was fixed. In order to check the model’s stability for estimating $J$ when estimaing the spatialy varied value of $D$,  the distribution of estimated values of $J$ is shown in Fig. 13. When $0.4 \leq D_\text{in} \leq 1.6$, the estimated values of $J$ were nearby the true value of 2.0. However, when $D_\text{in} = 0.1$ and $1.9$ in the \textbf{inner} region, the estimated value come to around 1.5, in spite the truth value was J = 2.0.

\section{Discussion}\label{discussion}
\subsection{Difficulty in Estimating $D$ and $J$}
As shown in Fig. 7 and Table II, the estimation error for $D$ was higher than that for $J$ in all models. Let us discuss the reasons for this observation. Looking at Fig. 3, the magnetic domains tend to become finer as $D$ and $J$ increase. For $J$, when $J = 0.2$ and $1.4 < D < 2.0$, there is one large island-shaped magnetic domain and a few small ones, while at $J=2.0$, numerous small island-shaped magnetic domains or labyrinthine domains appear, indicating a significant trend of finer magnetic domains. This pattern change is more pronounced for $J$ than for $D$, as can be seen in Fig. 3. On the other hand, for $D$, the labyrinthine structure observed at $D = 0.1$ and $1.8 < J < 2.0$ collapses as $D$ increases, resulting in numerous disordered island-shaped magnetic domains, but the degree of change is not as pronounced for $D$ as it is for $J$. From this subjective evaluation, it can be understood that estimating $D$ from magnetic domain patterns is more challenging than estimating $J$.

\subsection{Utility of Large-Scale Models}
The  \textbf{mod} model had lower estimation accuracy compared with \textbf{reg} and \textbf{reg\_pret} models; particularly the difference was pronounced in inference of $D$. As shown in Fig. 8,  \textbf{mod} was insensitive to changes in $D$, overestimating the value when $D$ was small ($< 1.3$) and underestimating it when $D$ was large ($> 1.3$). This can be attributed to the limited number of parameters in the  \textbf{mod} model, which prevented it from fully learning the complexity of the magnetic domain patterns (underfitting) and causing the model to estimate values close to the average value of $D$ in the training data, 1.38. This tendency was also observed in the \textbf{reg} and \textbf{reg\_pret} models, but to a lesser extent, demonstrating that large-scale models (\textbf{reg} and \textbf{reg\_pret}) can better handle the complexity of the task. This leads us to conclude that large-scale models can achieve higher estimation accuracy than small-scale models when learning the magnetic domain patterns addressed in this research.

\subsection{Utility of Pretraining}
The error of the \textbf{reg\_pret} model with pretraining was consistently smaller than that of \textbf{reg} model without pretraining. The improvement in accuracy was particularly large for $N_\text{data}=1$ and the target variable $D$ where it increased by 0.042, a larger improvement than when transitioning from the  \textbf{mod} model to the \textbf{reg} model (0.023). Even in the case with the smallest improvement, with $N_\text{data}=4$ and the target variable $J$, accuracy improved by 0.011. These results demonstrate the usefulness of pretraining with natural images for the analyzing magnetic domain patterns.

Moreover, when $N_\text{data}=1$, the accuracy improvement due to pretraining was 0.042 for $D$ and 0.014 for $J$, while for $N_\text{data}=4$, the respective values were 0.035 and 0.012. The accuracy improvements were more significant  when the data was limited, which is consistent with the empirical rule that pretraining is particularly effective when data is scarce.

\subsection{Comparison Based on the Amount of Training Data, $N_\text{data}$}
Both the \textbf{reg} and \textbf{reg\_pret} models improved in accuracy as the amount of training data increased. By comparison, the  \textbf{mod} model experienced a slight decrease in accuracy as the data volume increased. This numerical experimental result is consistent with the empirical rule that larger models benefit more from increased training data. The  \textbf{mod} model, in contrast, had very few parameters, and as a result, its error did not decrease even when it used more data than $N_\text{data} = 1$. The error in both $N_\text{data} = 4$ and $N_\text{data} = 1$ cases was nearly identical, while the slightly higher median error for $N_\text{data} = 4$ is believed to be attributable to random variations in the data.

\subsection{Causes of outliers}\label{Outlier_analysis}
Even the best model \textbf{reg\_pret} had some significant outliears as marked by the red square and diamond markers in top row, middle column pannel of Fig. 8. Fig.9 showcases the two patches of test data responsible for these outliers. We identified the two patches where with the true value of $D = 0.7$, the estimated value was 1.3 (Fig. 9(a)), and where with the true value of $D = 1.3$, the estimated value was 0.2 (Fig. 9(b)). The reason of the former outler is speculated that it is difficult to predict the parameter from the entire black patch. On the other hand, the reason of the latter is speculated that there exists the resemble learning patch whose true value of $D$ is 1.3 (Fig. 9(c)).
Because the $64 \times 64$ pixel patches cropped from the $256 \times 256$ pixel images are used for training and estimation, the cropped small patch might not contain the features significant for parameter estimation (Fig. 9(a)), and might be similar to a domain pattern generated by other parameters (Fig. 9(c)). Therefore, there is a possibility that our framework induces outliers by these factors. To overcome the issues, we need to increase the field of view of input image while we should keep the locality of estimation as described below.

\subsection{Challenges in Detecting Spatial Variations of Parameters}
As demonstrated in Sect.\ref{Detection of Spatial Variations in Parameters from Magnetic Domain Patterns}, our method can detect changes in simulation parameters from the magnetic domain patterns. As shown in the ROC analysis of the estimated value distribution in Fig. 12, even when the difference between the \textbf{inner} and \textbf{outer} $D$ values was 0.3, the AUC score was 0.798. According to the conventional criterion that indicates a classifier as good if its AUC score is above 0.75, we can conclude that our method is capable of detecting this change. However, the estimated values have an interquartile range of approximately 0.3, making it difficult to detect parameter changes with a resolution below this value. Thus, the accuracy still has to be improved in order to detect parameter changes at a higher resolution. 

As described in Sect IIIB, when estimaing the spatialy varied value of $D$, the estimation for other parameter $J$ was influenced by the great change in $D_\text{in}$ (at $D_\text{in} \in \{0.1, 1.9\}$). Therefore, we need to improve the estimation stability of target parameters to be independent of values of other parameters.

As discussed in Sect IV E, to realize the sliding window approach, the field of view of the input image to the models was limited to $64 \times 64$ pixel small region, which resulted in the outliers. Thus, we need to reduce the outliers caused by limited field-of-view.

To overcome the above three issues in estimating spatial varying parameters, we need to increase the field of view of input image while we should keep the locality of estimation. The fully-convolutional model\cite{7298965}, which can perform the pixel-wise segmentation or regression from the full size input image, can be used for achieving this purpose. The fully-convolutional model offers a benefit of no requirement to crop small patches from the whole image, unlike the sliding window approach, allowing pixel-wise learning and inference on the full-size images. Thereore, the fully-convolutional model might detect parameter changes at higher resolution, and might be less likely to detect the outliers. Moreover, this model might stably estimate target parameters independent of other parameter changes due to wide field-of-view of input image. In addition, while the sliding window approach must repeat inference as many time as the number of pixels within the image, the fully-convolutional model estimates parameters across the entire image at once, providing a computational time advantage.

Additionally, even when the simulation model parameters change discontinuously, magnetic domain patterns continuously change due to the exchange and dipole interactions. Thus, as shown in Fig. 11, the parameter estimation from continuous changing patterns around the boundary is challenging. To improve the estimation ability around the boundary, Bayesian inference, which outputs both the estimate of the parameter and its confidence simultaneously or segmentation models\ref{rs13193836} that estimate the boundary of parameter changes can be applied.

\subsection{Applicability of the Proposed Method to Real-World Data}
The parameters estimated in this study are the material noise parameter $D$ and the macroscale exchange interaction parameter $J$. The noise parameter represents defects that emerge during material fabrication, particularly those leading to changes in magnetic anisotropy. In screening actual materials, it might be possible to identify fabrication conditions that result in fewer defects by investigating conditions that produce materials with minimal defects. Materials with strong exchange interactions tend to exhibit ferromagnetic properties, and the exchange interaction plays a role in determining the Curie temperature\cite{article_coe}, which relates to the material's practicality. While strong exchange interactions are desirable for permanent magnets, weaker exchange interactions are more suitable for applications such as optical isolators. At the atomic level, the exchange interaction is determined by the type and combination of elements in the material\cite{doi:10.1021/ic961448x}. However, to the best of our knowledge, no method has been established for designing exchange interactions at larger, macroscale levels.

The simulations used in this study employed a sufficiently large length scale in order to disregard the atomic structure of the material. If the proposed method can estimate the exchange interaction on the macroscopic scale, it may potentially contribute to the development of materials with exchange interaction parameters suitable for specific applications.

The measurement of the exchange interaction in magnetic thin films is frequently conducted through ferromagnetic resonance spectroscopy and Brillouin light scattering. These techniques, while comprehensive, are time-consuming for the measurement process itself, and moreover are necessitating additional computational time for postexperimental data analysis. On the other hand, our proposed method is assumed to be applied to images acquired with Kerr effect microscopy which allows rapid capturing of magnetic thin-film domain patterns. Due to the rapidity of the imaging with the microscopy and the inference with our proposed method, it facilitates the expeditious estimation of thin film physical parameters including the exchange interaction, compared to the conventional methods described above. Espeticially, the advantage becomes significantly pronounced if our method is applied to capture parameters more difficult to measure such as the Dzyaloshinskii-Moriya interaction. While the procurement of simulation data and the training of the models with the procured data are time-intensive endeavors, the sharing of the trained models can significantly reduce the introduction cost of our method.

\section{Conclusion}\label{conclusion}
This study aimed to estimate the spatial distribution of non-uniform physical parameters by using CNNs for the analysis of polycrystalline thin films. For all magnetic domain patterns, the physical parameters were estimated from the patterns within small subregions within a window, and the spatial distribution of physical parameters was estimated by shifting this window. To improve the accuracy of the parameter estimation in these small subregions, we demonstrated the effectiveness of large-scale models used in natural image classification and the usefulness of pretraining. Using a model with improved estimation accuracy in small subregions (\textbf{reg\_pret}), we performed inference on simulation data with spatially varying parameters and demonstrated the detection of parameter changes.

This finding suggests the potential for determining thin-film characteristics from small subregion magnetic domain patterns and detecting spatial changes in thin-film properties. By further developing this research, it could advance magnetic thin film materials. Future work should involve applying this method to more realistic polycrystalline simulation data and real material data. At the same time, it will be necessary to improve the accuracy of parameter estimation and address the interpretability issues common to CNNs.

The proposed method is not limited to the analysis of magnetic domain patterns in the hysteresis process of magnetic thin films. Thus, we can also expect this method to be useful for analyzing patterns in other phenomena, like polycrystalline growth in metals.

\section{Appendix}\label{appendix}
\subsection{Combinations of $D$ and $J$ used in the dataset}
Some combinations of $D$ and $J$ resulted in diverging calculations, making it impossible to perform hysteresis process simulations. We excluded such combinations from the dataset. The combinations of $D$ and $J$ used in the dataset are shown in Table III.

\begin{table}[h]
\centering
\caption{\textbf{Combinations of $D$ and $J$ in our dataset}}
\label{table:dj_combinations}
\begin{tabular}{c@{\hskip 20pt}c@{\hskip 10pt}||@{\hskip 10pt}c@{\hskip 20pt}c}
$J$ & $D$ & $J$ & $D$ \\
\hline
2.0 & 0.1, ..., 2.0 & 1.0 & 0.9, ..., 2.0 \\
1.9 & 0.1, ..., 2.0 & 0.9 & 1.2, ..., 2.0 \\
1.8 & 0.1, ..., 2.0 & 0.8 & 1.3, ..., 2.0 \\
1.7 & 0.1, ..., 2.0 & 0.7 & 1.4, ..., 2.0 \\
1.6 & 0.1, ..., 2.0 & 0.6 & 1.4, ..., 2.0 \\
1.5 & 0.1, ..., 2.0 & 0.5 & 1.6, ..., 2.0 \\
1.4 & 0.1, ..., 2.0 & 0.4 & 1.7, ..., 2.0 \\
1.3 & 0.1, ..., 2.0 & 0.3 & 1.8, ..., 2.0 \\
1.2 & 0.5, ..., 2.0 & 0.2 & 1.8, ..., 2.0 \\
1.1 & 0.6, ..., 2.0 & 0.1 & 2.0 \\
\end{tabular}
\end{table}

\subsection{Accuracies of various deep learning models}
\begin{figure*}[t]\label{fig:manymodels_boxplot}
\centering
\includegraphics[width=\linewidth]{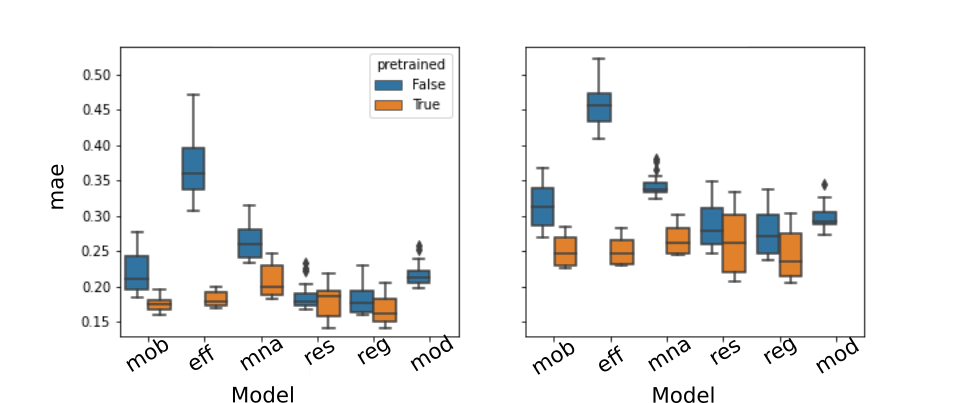}
\caption{\textbf{Accuracy comparison of various learning models}\\ For spatially uniform data, we performed 20 rounds of training and inference using the same method as described in Sect.\ref{numerical experiment}. The mean absolute errors are presented them in a box-and-whisker plot.}
\end{figure*}

In addition to the detailed Model0-6(\textbf{mod}) and RegNet(\textbf{reg}), we also conducted numerical experiments on MobilenetV3-Large(\textbf{Mob}), EfficientNet-B3(\textbf{eff}), MnasNet1-0(\textbf{mna}), and ResNext-50(\textbf{res}). ResNext is a model that incorporates residual blocks to address the vanishing gradient problem\cite{He2016DeepRL} and split-transform-merge\cite{7298594} that branches the input features of the residual blocks, transforms them individually, and then merges them. MnasNet is a model designed to achieve both a fast computation and high recognition accuracy on mobile devices such as smartphones. It incorporating residual blocks, split-transform-merge, and squeeze-and-excitation\cite{DBLP:journals/corr/abs-1709-01507}, and it learns and infers weights for each feature channel. Here, reinforcement learning methods are used to achieve good inference accuary. MobilenetV3 is based on MnasNet. It achieves speed and accuracy improvements by performing a more granular parameter search by using NetAdapt\cite{DBLP:journals/corr/abs-1804-03230}. EfficientNet-B3 is also based on MnasNet and has achieved efficient and high recognition accuracy models by carefully examining the relationship between accuracy and model scaling. Regnet explores models in a higher degree of freedom search space than these models and has improved accuracy over EfficientNet for large-scale models.

We trained and performed inference on these models on spatially uniform data following the same procedure as in Sect.\ref{training}. The results are shown in Fig. 14. The median error for the res model's $J$ inference was slightly worse at 0.004 when it was pretrained, but in all other cases, the pretrained models performed better. Furthermore, all pretrained models had higher accuracy than the small-scale  \textbf{mod} model, supporting the conclusions of this paper on the usefulness of large-scale models and pretraining.

\bibliographystyle{unsrt}
\bibliography{apssamp}
\end{document}